\pdfoutput=1
\documentclass[prl,twocolumn,showpacs,amsmath,amssymb,superscriptaddress,floatfix]{revtex4}

\usepackage{graphicx}
\usepackage{dcolumn}
\usepackage{bm}
\usepackage{epsfig}

\begin{document}

\title{Multiterminal single-molecule--graphene-nanoribbon thermoelectric devices with gate-voltage tunable figure of merit $ZT$}

\author{Kamal K. Saha}
\affiliation{Department of Physics and Astronomy, University of Delaware, Newark, DE 19716-2570, USA}
\author{Troels Markussen}
\affiliation{Center for Atomic-scale Materials Design (CAMD), Department of Physics, Technical University of Denmark, DK-2800 Kongens Lyngby, Denmark}
\author{Kristian S. Thygesen}
\affiliation{Center for Atomic-scale Materials Design (CAMD), Department of Physics, Technical University of Denmark, DK-2800 Kongens Lyngby, Denmark}
\author{Branislav K. Nikoli\' c}
\email{bnikolic@udel.edu}
\affiliation{Department of Physics and Astronomy, University of Delaware, Newark, DE 19716-2570, USA}

\begin{abstract}
We study thermoelectric devices where a single \mbox{18-annulene} molecule is connected to metallic zigzag graphene nanoribbons (ZGNR) via highly transparent contacts that allow for injection of evanescent wave functions from ZGNRs into the molecular ring.  Their overlap generates a peak in the electronic transmission, while ZGNRs additionally suppress hole-like contributions to the thermopower. Thus optimized thermopower, together with suppression of phonon transport through ZGNR-molecule-ZGNR structure, yield the thermoelectric figure of merit $ZT \sim 0.5$ at room temperature and $0.5 < ZT < 2.5$ below liquid nitrogen temperature. Using the nonequilibrium Green function formalism combined with density functional theory, recently extended to multiterminal devices, we show how the transmission resonance can also be manipulated by the voltage applied to a third ZGNR electrode, acting as the top gate covering molecular ring, to tune the value of $ZT$.
\end{abstract}

\maketitle

Thermoelectrics transform temperature gradients into electric voltage and vice versa. Although a plethora of thermoelectric-based energy conversion and cooling applications has been envisioned, their usage is presently limited by their small efficiency~\cite{Vining2009}. Careful tradeoffs are needed to optimize the dimensionless figure of merit $ZT=S^2GT/\kappa$ quantifying the maximum efficiency of a thermoelectric conversion because $ZT$ contains unfavorable combination of the thermopower $S$, average temperature $T$, electrical conductance $G$ and the total thermal conductance $\kappa =\kappa_{\rm el} + \kappa_{\rm ph}$ (including   contributions from electrons $\kappa_{\rm el}$ and phonons $\kappa_{\rm ph}$). The devices with $ZT > 1$ are regarded as good thermoelectrics, but $ZT > 3$ is required to compete with conventional generators~\cite{Vining2009}.

The major experimental efforts to increase $ZT$ have been focused on suppressing the phonon conductivity using either complex (through disorder in the unit cell) bulk materials~\cite{Snyder2008} or bulk nanostructured materials~\cite{Minnich2009}. A complementary approach engineers  electronic density of states to obtain a sharp singularity~\cite{Minnich2009} near the Fermi energy  which can enhance the power factor $S^2G$.

The very recent experiments~\cite{Reddy2007} and theoretical studies~\cite{Paulsson2003,Ke2009,Nozaki2010,Entin-Wohlman2010,Murphy2008} have ignited exploration of devices where a single molecule is attached to metallic~\cite{Ke2009} or semiconducting~\cite{Nozaki2010} electrodes, so that dimensionality reduction and possible strong electronic correlations~\cite{Murphy2008} allow to increase $S$ concurrently with diminishing $\kappa_{\rm ph}$ while keeping the nanodevice disorder-free~\cite{Markussen2009}. For example, creation of sharp transmission resonances near the Fermi energy $E_F$ by tuning the {\em chemical properties} of the molecule and molecule-electrode contact can substantially enhance the thermopower $S$ which depends on the derivative of the conductance near $E_F$. At the same time, the presence of a molecule in the electrode-molecule-electrode heterojunction severely disrupts phonon propagation when compared to homogenous clean electrode.

\begin{figure}
\includegraphics[scale=0.33,angle=0]{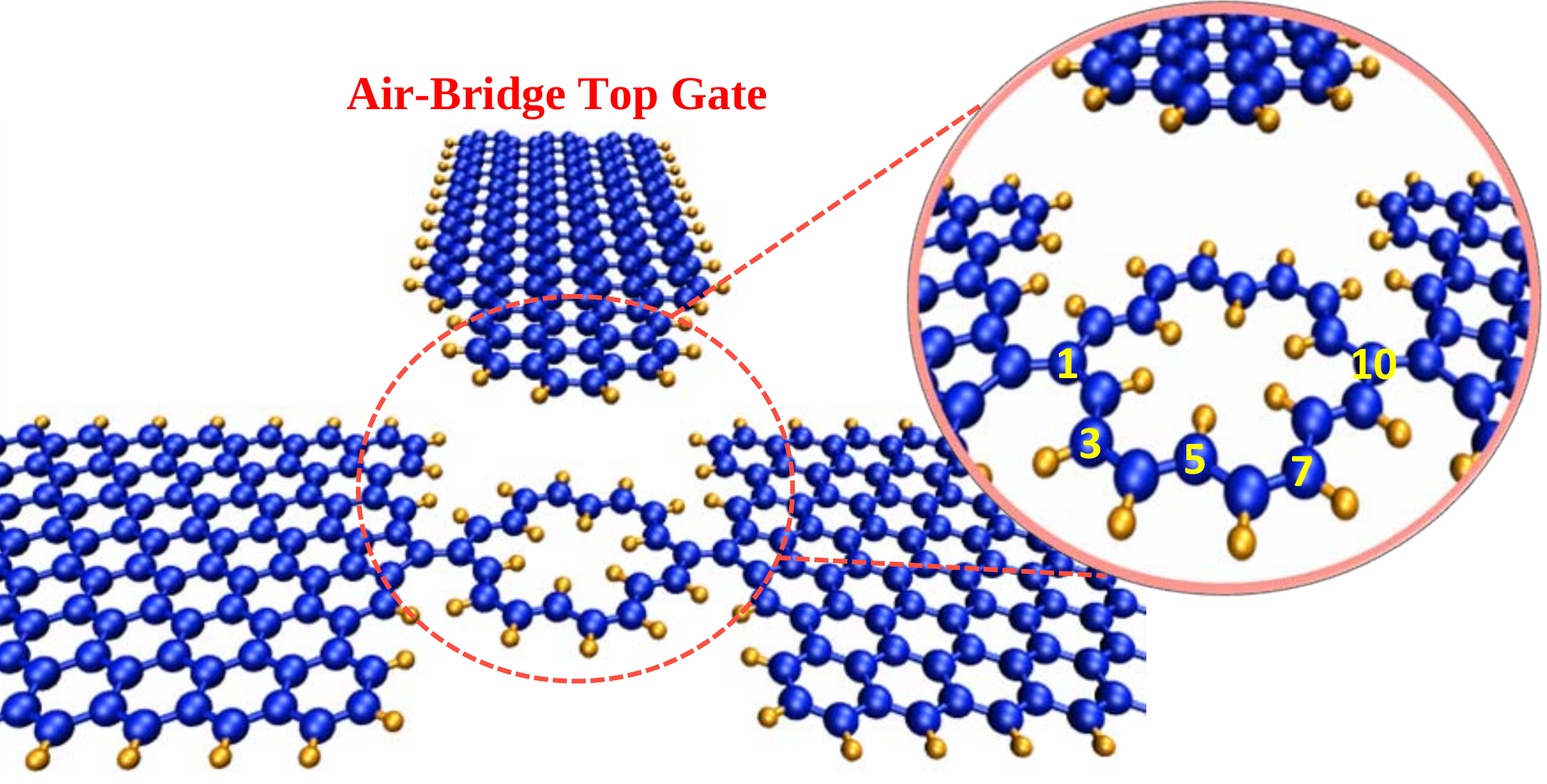}
\caption{(Color online) Schematic view of the proposed ZGNR$|$18-annulene$|$ZGNR three-terminal heterojunction. The contact between the source and drain 8-ZGNR (consisting of 8 zigzag chains) metallic electrodes and a ring-shaped  molecule is made via 5-membered rings of carbon atoms (dark blue), while the electrodes are attached to atoms 1 and 10 of the molecule. The third electrode is coupled as an ``air-bridge'' top gate, made of ZGNR as well, covering only molecular ring  at the distance \mbox{5.3 \mbox{\AA}}. The two-terminal version of the device assumes that such top gate is absent. The hydrogen atoms (light yellow) are included to passivate the edge carbon atoms.}
\label{fig:setup}
\end{figure}

\begin{figure}
\includegraphics[scale=0.33,angle=0]{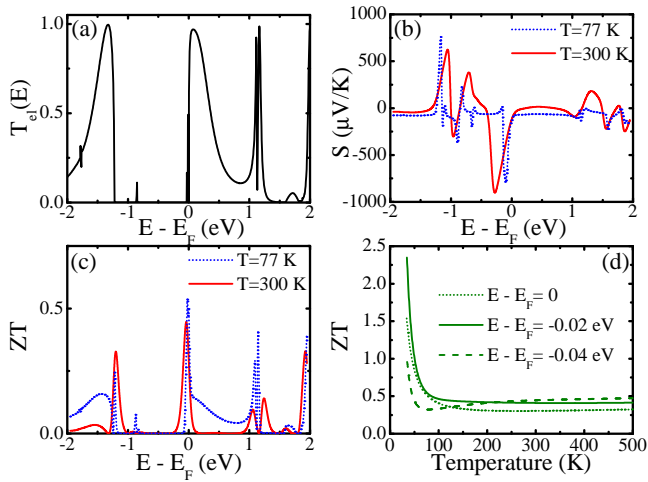}
\caption{(Color online) Physical quantities determining thermoelectric properties of the two-terminal device  shown in Fig.~\ref{fig:setup} (without the top gate electrode): (a) zero-bias electronic transmission $\mathcal{T}_{\rm el}(E)$; (b) the corresponding thermopower $S$ at two different temperatures; (c) thermoelectric figure of merit $ZT$ vs. energy at two different temperatures; and (d) $ZT$ vs. temperature at three different energies.}
\label{fig:fig2}
\end{figure}

\begin{figure}
\includegraphics[scale=0.2,angle=0]{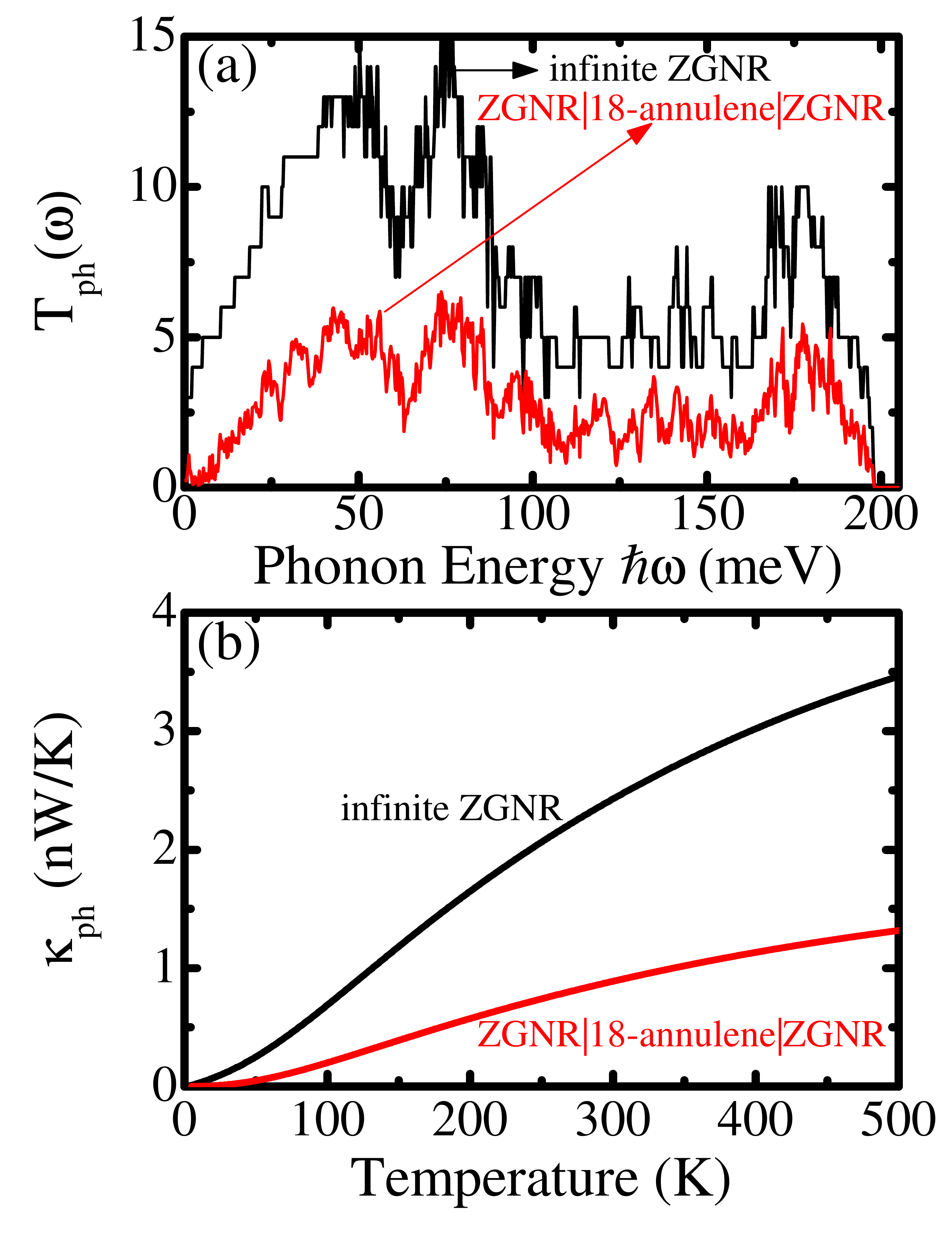}
\caption{(Color online) (a) The phonon transmission function $\mathcal{T}_{\rm ph}(\omega)$ and (b) the corresponding thermal conductance $\kappa_{\rm ph}$ for an infinite 8-ZGNR and the two-terminal device shown in Fig.~\ref{fig:setup} (assuming absence of the third top gate electrode) whose source and drain electrodes are  made of semi-infinite 8-ZGNR.}
\label{fig:fig3}
\end{figure}

\begin{figure}
\includegraphics[scale=0.33,angle=0]{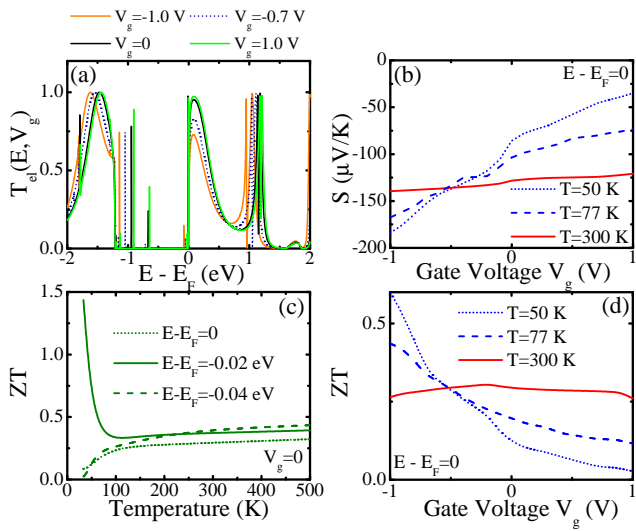}
\caption{(Color online) Physical quantities determining thermoelectric properties of the three-terminal device shown in Fig.~\ref{fig:setup} as a function of the applied gate voltage $V_g$: (a) zero-bias electronic transmission $\mathcal{T}_{\rm el}(E,V_g)$; (c) the corresponding thermopower $S$ at two different temperatures; (c) thermoelectric figure of merit $ZT$ vs. temperature at $V_g=0$; and (d) $ZT$ vs. $V_g$ at two different temperatures.}
\label{fig:fig4}
\end{figure}

In this Letter, we exploit a transparent contact between metallic zigzag graphene nanoribbon (ZGNR) electrodes and a ring-shaped \mbox{18-annulene} molecule to propose two-terminal (i.e., top gate absent in Fig.~\ref{fig:setup}) and three-terminal devices illustrated in Fig.~\ref{fig:setup}.  Their  thermoelectric properties  are analyzed using the very recently developed nonequilibrium Green function combined with density functional theory  formalism for multiterminal nanostructures (MT-NEGF-DFT)~\cite{Saha2009a,Saha2010}. The high contact transparency allows evanescent modes from the two ZGNR electrodes to tunnel into the molecular region and meet in the middle of it (when the molecule is short enough~\cite{Ke2007}), which is a counterpart of  the well-known metal induced gap states  in metal-semiconductor Schottky junctions. This effect can induce a large peak (i.e., a resonance) in the electronic transmission function near $E_F$ [Fig.~\ref{fig:fig2}(a)], despite the HOMO-LUMO energy gap of the isolated molecule. The enhancement of the thermopower [Fig.~\ref{fig:fig2}(b)] due to transmission resonance around $E \ge E_F$ and suppression of hole-like contribution (i.e., negligible transmission around $E<E_F$) to $S$, together with several times smaller phonon thermal conductance   (Fig.~\ref{fig:fig3}) when compared to infinite ZGNR, yields the maximum room-temperature $ZT \sim 0.5$ in the two-terminal device [Fig.~\ref{fig:fig2}(c),(d)]. Furthermore, we discuss how a third top gate ZGNR electrode covering the molecule, while being separated  by an air gap in Fig.~\ref{fig:setup},  can tune the properties of the transmission resonance via the applied gate voltage thereby making possible further enhancement of $ZT$ (Fig.~\ref{fig:fig4}).

Among the recent theoretical studies of molecular thermoelectric devices via the NEGF-DFT framework~\cite{Ke2009,Nozaki2010}, most have been focused~\cite{Ke2009} on computing the thermopower $S$, with only a few~\cite{Nozaki2010} utilizing DFT to obtain forces on displaced atoms and then compute $\kappa_{\rm ph}$. Moreover, due the lack of NEGF-DFT algorithms for {\em multiterminal} nanostructures, the possibility to tune thermoelectric properties of  single-molecule devices via the usage of the third electrode has remained largely an unexplored realm~\cite{Entin-Wohlman2010}. We note that the recent proposal~\cite{Nozaki2010} for the two-terminal molecular thermoelectric devices with sophisticated combination of a local chemical tuning of the molecular states and usage of semiconducting electrodes has predicted much smaller $ZT \sim 0.1$ at room temperature. In addition, our $0.5<ZT<2.5$ at $E-E_F=-0.02$ eV (which can be set by the backgate electrode covering the whole device~\cite{Zuev2009}) in the temperature range \mbox{$T = $ 30--77 K} is much larger than the value achieved in conventional low-temperature bulk thermoelectric materials~\cite{Snyder2008}.

The recent fabrication of GNRs with ultrasmooth edges~\cite{Cai2010}, where those with zigzag edges are insulating at very low temperatures due to edge magnetic ordering which is nevertheless easily destroyed above $\gtrsim 10$ K~\cite{Yazyev2008}, has opened
new avenues for highly controllable molecular junctions with a  well-defined molecule-electrode contact characterized by  high transparency, strong directionality and reproducibility. This is due to the fact that strong molecule-GNR \mbox{$\pi$-$\pi$} coupling makes possible  formation of a continuous \mbox{$\pi$-bonded} network across GNR and orbitals of conjugated organic molecules~\cite{Ke2007}. Unlike the metallic carbon nanotubes (CNTs) employed experimentally~\cite{Guo2006} to generate such networks~\cite{Ke2007},  GNRs have planar structure appropriate for aligning and patterning.

The early experiments~\cite{Guo2006} on CNT$|$molecule$|$CNT heterojunctions  have measured surprisingly small conductances for a variety of  sandwiched molecules. The first-principles analysis of different setups reveals that this is due to significant twisting forces when molecule is connected to CNT via, e.g., 6-membered rings~\cite{Ke2007}. Therefore, to keep nearly parallel and in-plane configuration (hydrogen atoms of \mbox{18-annulene} slightly deviate from the molecular plane) of our ZGNR$|$18-annulene$|$ZGNR junction, we use a 5-membered ring~\cite{Ke2007} in Fig.~\ref{fig:setup} to chemically bond ZGNR to annulene.

The carbon atoms of a ring-shaped \mbox{18-annulene} molecule can be connected to ZGNR electrodes in configurations whose Feynman paths for electrons traveling around the ring generate either constructive or destructive quantum interference effects imprinted on the conductance~\cite{Markussen2010}. For example, a $\pi$-electron at $E_F$ entering the molecule in setup (1,10) shown in Fig.~\ref{fig:setup} has the wavelength $k_F/2d$ ($d$ is the spacing between carbon atoms within the molecule), so that for the two simplest Feynman paths of length $9d$ (upper half of the ring) and $9d$ (lower half of the ring) the phase difference is 0.  Note that the destructive quantum interference~\cite{Markussen2010} would form an additional dip (i.e., antiresonance) within the main transmission peak in Fig.~\ref{fig:fig2}(a)  generated by the injection of overlapped evanescent modes~\cite{Saha2010}. The effect of antiresonance on the thermopower $S$ for gold$|$18-annulene$|$gold junctions has been studied in Ref.~\cite{Bergfield2009} as a possible sensitive tool to confirm the effects of quantum coherence on transport through single-molecule junctions.

The computation of quantities entering $ZT$ for realistic single-molecule junctions requires quantum transport methods combined with {\em first-principles} input about atomistic and electronic structure to capture {\em charge transfer} in equilibrium (which is indispensable to obtain correct zero-bias transmission of, e.g.,  carbon-hydrogen systems~\cite{Areshkin2010}), geometrically-optimized atomic positions of the molecular bridge including molecule-electrode separation in equilibrium, and forces on atoms when they are perturbed out of equilibrium. The state-of-the-art approach that can capture these effects, as long as the coupling between the molecule and the electrodes is strong enough to ensure transparent contact and diminish Coulomb blockade effects~\cite{Murphy2008,Cuniberti2005}, is NEGF-DFT~\cite{Cuniberti2005,Areshkin2010}.

The technical details of the construction of the nonequilibrium density matrix via NEGF-DFT for multiterminal devices are discussed in Ref.~\cite{Saha2009a}. Our MT-NEGF-DFT code utilizes  ultrasoft pseudopotentials and Perdew-Burke-Ernzerhof (PBE) exchange-correlation functional. The localized basis set for DFT calculations  is constructed from atom-centered orbitals (six per C atom and four per H atom)  that are optimized variationally for the electrodes and the central molecule separately while their electronic structure is obtained concurrently.

In the coherent transport regime, the NEGF post-processing of the result of the DFT self-consistent loop expresses the zero-bias electron transmission function between the left (L) and the right (R) electrodes as:
\begin{equation}\label{eq:telectron}
\mathcal{T}_{\rm el}(E) = {\rm Tr} \left\{ {\bm \Gamma}_L (E)  {\bf G}(E) {\bm \Gamma}_R (E)  {\bf G}^\dagger(E)  \right\}.
\end{equation}
The matrices \mbox{${\bm \Gamma}_{L,R}(E)=i[{\bm \Sigma}_{L,R}(E) - {\bm \Sigma}_{L,R}^\dagger(E]$} account for the level broadening due to the coupling to the electrodes, where ${\bm \Sigma}_{L,R}(E)$ are the self-energies introduced by the ZGNR electrodes~\cite{Areshkin2010}. The retarded Green function matrix is given by ${\bf G}=[E{\bf S} - {\bf H} - {\bm \Sigma}_L - {\bm \Sigma}_R]^{-1}$, where in the local orbital basis $\{ \phi_i \}$ Hamiltonian matrix ${\bf H}$ is composed of elements $H_{ij} = \langle \phi_i |\hat{H}_{\rm KS}| \phi_{j} \rangle$ ($\hat{H}_{\rm KS}$ is the effective Kohn-Scham Hamiltonian obtained from the DFT self-consistent loop) and the overlap matrix ${\bf S}$ has elements $S_{ij} = \langle \phi_i | \phi_j \rangle$.

The transmission function Eq.~(\ref{eq:telectron}) obtained within the NEGF-DFT framework allows us to compute the following integrals~\cite{Esfarjani2006}
\begin{equation}\label{eq:kintegral}
K_n(\mu) = \frac{2}{h} \int\limits_{-\infty}^{\infty} dE\, \mathcal{T}_{\rm el}(E)  (E - \mu)^n \left(-\frac{\partial f(E,\mu)}{\partial E} \right),
\end{equation}
where \mbox{$f(E,\mu)=\{ 1 + \exp[(E-\mu)/k_BT] \}^{-1}$} is the Fermi-Dirac distribution function at the chemical potential $\mu$. The knowledge of $K_n(\mu)$ finally yields all electronic quantities in the expression for $ZT$: $G=e^2K_0(\mu)$; $S=K_1(\mu)/[eTK_0(\mu)]$; and $\kappa_{\rm el} = \{K_2(\mu) - [K_1(\mu)]^2/K_0(\mu)\}/T$.

The phonon thermal conductance is obtained from the phonon transmission function $\mathcal{T}_{\rm ph}(\omega)$ using the corresponding Landauer-type formula for the scattering region (molecule + portion of electrodes) attached to two semi-infinite electrodes:
\begin{equation}\label{eq:tphonon}
\kappa_{\rm ph} = \frac{\hbar^2}{2\pi k_B T^2} \int\limits_{0}^{\infty} d\omega\, \omega^2 \mathcal{T}_{\rm ph}(\omega) \frac{ e^{\hbar\omega/k_BT}}{(e^{\hbar\omega/k_BT}-1)^2}.
\end{equation}
The phonon transmission function $\mathcal{T}_{\rm ph} (\omega)$ can be
calculated from the same Eq.~(\ref{eq:telectron}) with substitution
${\bf H} \rightarrow {\bf K}$ and ${\bf S} \rightarrow \omega^2 {\bf
M}$, where ${\bf K}$ is the force constant matrix and ${\bf M}$ is a
diagonal matrix with the atomic masses. We obtain the force constant
matrix using GPAW, which is a real space electronic structure code based
on the projector augmented wave method~\cite{Enkovaara2010}. The
electronic wave functions are expanded in atomic orbitals with a
single-zeta polarized basis set, and PBE exchange-correlation
functional is used. The whole scattering region, which includes 27 layers of ZGNR electrodes,
is first relaxed to a maximum force of $0.01$ eV/\AA  per atom. Subsequently, we displace each atom $I$ by $Q_{I\alpha}$
in direction $\alpha=\{x,y,z\}$ to get the forces $F_{I\alpha,J\beta}$
on atom $J\neq I$ in direction $\beta$. The elements of ${\bf K}$ are
then computed from finite differences,
${\bf K}_{I\alpha,J\beta} =
[F_{J\beta}(Q_{I\alpha})-F_{J\beta}(-Q_{I\alpha})]/2Q_{I\alpha}$. The
intra-atomic elements are
calculated by imposing momentum conservation, such that
$K_{I\alpha,I\beta} = -\Sigma_{J\neq I}K_{I\alpha,J\beta}$.

Figure~\ref{fig:fig2}(a) shows the zero-bias electronic transmission $\mathcal{T}_{\rm el}(E)$ for the two-terminal version of the device in Fig.~\ref{fig:setup}, where the peak near $E_F$ is conspicuous. Additionally, the suppression of the hole-like transmission [$\mathcal{T}_{\rm el}(E) \rightarrow 0$ around $E<E_F$] evades  unfavorable compensation~\cite{Nozaki2010} of hole-like and electron-like contributions to the thermopower. This is due to the symmetry of the valence band  propagating transverse mode in GNR semi-infinite electrode which changes  sign at the two carbon atoms closest to the molecule for the geometry (molecule connected to the middle of the GNR edge) in Fig.~\ref{fig:setup}. We emphasize that these features of $\mathcal{T}_{\rm el}(E)$ are quite insensitive to the details of (short)  conjugated molecules, and since they are governed by the ZGNR Bloch states, they are {\em impervious} to the usual poor estimates of the band gap size and molecular energy level position in DFT.

The maximum value of the corresponding thermopower $S$ plotted in Fig.~\ref{fig:fig2}(b) is slightly away from $E-E_F=0$ and it is an {\em order of magnitude larger} than the one measured on large-area  graphene~\cite{Zuev2009} or in molecular junctions with gold electrodes~\cite{Reddy2007}. Moreover, the interruption of the infinite ZGNR by a molecule acts unfavorably to phonon transmission $\mathcal{T}_{\rm ph}(\omega)$, thereby generating three times smaller $\kappa_{\rm ph}$ at room temperature when compared in Fig.~\ref{fig:fig3} to the thermal conductance of an infinite 8-ZGNR. Figures~\ref{fig:fig2}(c),(d) demonstrate that the interplay of large $S^2G$ and reduced $\kappa_{\rm ph}$ for the proposed two-terminal device yields the room-temperature $ZT \sim 0.5$ around $E-E_F=0$.

The introduction of the narrow air-bridge top gate electrode in Fig.~\ref{fig:setup}, which is positioned at the distance \mbox{5.3 \mbox{\AA}} (ensuring negligible tunneling leakage current into such third ZGNR electrode) away from the two-terminal device underneath while covering only the molecular ring, makes possible tuning of the transmission resonance shown in Fig.~\ref{fig:fig4}(a). Even in the absence of any applied gate voltage ($V_g=0$), $ZT$ vs. temperature plotted in Fig.~\ref{fig:fig4}(c) is notably modified when compared to the corresponding functions in Fig.~\ref{fig:fig2}(d) for the two-terminal device. This stems from slight hybridization of the top gate and molecular states.  The narrowing of the transmission peak around $E_F$ due to the application of negative gate voltage enhances the thermopower in Fig.~\ref{fig:fig4}(b), thereby increasing $ZT$ above its value at $V_g=0$, which can be substantial at low temperatures as shown in Fig.~\ref{fig:fig4}(d).

In conclusion, we predict that a single conjugated molecule attached to metallic GNR electrodes, where the  transparent molecule-GNR contact allows evanescent modes to penetrate from the electrodes into the HOMO-LUMO molecular gap generating a transmission resonance, can act as an efficient thermoelectric device. Our first-principles  simulations suggest that its figure of merit can reach $ZT \sim 0.5$  at room temperature or $0.5<ZT <2.5$ below liquid nitrogen temperature, which is much higher than $ZT$ found in other recent proposals for molecular thermoelectric devices~\cite{Nozaki2010}. Moreover, introduction of the third air-bridge top gate covering the molecule can change the sharpness of the transmission resonance via the application of the gate voltage, thereby  opening a path toward further optimization of $ZT$. We anticipate that much higher $ZT$ could be achieved by testing different types of molecules to reduce $\kappa_{\rm ph}$ since the power factor $S^2G$ is already optimized in our device by the usage of GNR electrodes which generate molecular-level-independent transmission resonance while the symmetry of propagating modes in GNRs lifts the compensation of hole-like and electron-like contributions to $S$.

\begin{acknowledgments}
We thank K. Esfarjani for illuminating discussions. Financial support under NSF Grant No. ECCS 0725566 (K. K. S. and B. K. N.) and FTP Grant No. 274-08-0408 (T. M. and K. S. T.) is gratefully acknowledged. The supercomputing time was provided in part by the NSF through TeraGrid resource TACC Ranger under Grant No. TG-DMR100002. K. K. S. and B. K. N. thank CAMD at the Technical University of Denmark for their kind hospitality during which part of this work was done.
\end{acknowledgments}





\end{document}